\documentclass[conference]{IEEEtran}
\usepackage{amssymb}
\usepackage{amsthm}
\theoremstyle{remark}
\usepackage{booktabs}
\usepackage{multirow}

\usepackage{enumitem}
\usepackage{lipsum}

\makeatletter
\def\input@path{{./tables/}}
\makeatother

\usepackage{cite}

\ifCLASSINFOpdf
 \usepackage[pdftex]{graphicx}
 \graphicspath{{./figures/}}
\else
\usepackage[dvips]{graphicx}
\graphicspath{{./figures/}}
\fi
\usepackage{amsmath}
\ifCLASSOPTIONcompsoc
  \usepackage[caption=false,font=normalsize,labelfont=sf,textfont=sf]{subfig}
\else
  \usepackage[caption=false,font=footnotesize]{subfig}
\fi
\usepackage{url}

\IEEEoverridecommandlockouts
\begin{document}
%

\title{Hybrid European MV--LV Network Models\\
for Smart Distribution Network Modelling
\thanks{
This work was funded by the Engineering and Physical Sciences Research Council through grant no.  EP/S00078X/1 (Supergen  Energy Networks Hub 2018). Email: matthew.deakin@newcastle.ac.uk.
}
}


\author{
\IEEEauthorblockN{Matthew Deakin, \textit{Member, IEEE}, David Greenwood, \textit{Member, IEEE},\\ Sara Walker, \textit{Senior Member, IEEE}, Phil C. Taylor$^{\dagger}$, \textit{Senior Member, IEEE}}
\IEEEauthorblockA{School of Engineering, Newcastle University, Newcastle-upon-Tyne, UK\\
$^{\dagger}$Department of Electrical and Electronic Engineering, University of Bristol, Bristol, UK}
}

%
%

\maketitle

\begin{abstract}
A pair of European-style, integrated MV--LV circuits are presented, created by combining generic MV and real LV networks. The two models have 86,000 and 113,000 nodes, and are made readily available for download in the OpenDSS file format. Primary substation tap change controls and MV--LV feeders are represented as three-phase unbalanced distribution network models, capturing the coupling of voltages at the MV level. The assumptions made in constructing the models are outlined, including a preconditioning step that reduces the number of nodes by more than five times without affecting the solution. Two flexibility-based case studies are presented, with \mbox{TSO--DSO} and peer--peer-based smart controls considered. The demonstration of the heterogeneous nature of these systems is corroborated by the analysis of measured LV voltage data. The models are intended to aid the development of algorithms for maximising the benefits of smart devices within the context of whole energy systems.
\end{abstract}


%
\IEEEpeerreviewmaketitle

\section{Introduction}
\IEEEPARstart{T}{he} release of the IEEE 8500 Node feeder in 2010 \cite{dugan2010ieee} coincided with an unprecedented increase in smart control system development for distribution network operations. Robust test-beds such as this are necessary so that these controls can designed to scale efficiently to the sizes required for implementation by distribution system operators (DSOs) or aggregators \cite{givisiez2020review}. Smart controls are often designed to maximise the benefits of low-carbon Distributed Energy Resources (DERs), which can affect networks in both positive and negative senses. Domestic-scale DERs (such as heat pumps and electric vehicles) often have large power and energy requirements, and have therefore been proposed to provide valuable flexibility services at scales well above the LV level at which they are connected \cite{givisiez2020review}. 

Whilst the radial operation of distribution networks is almost ubiquitous in most localities (with the exception of dense urban areas), the size and number of customers fed by LV circuits is very different in European-- and North American--style circuits. The former is characterised by the use of extensive LV circuits (with hundreds of customers fed from a single LV transformer). Conversely, North American-style circuits have more extensive MV feeders and smaller LV sections, usually with just a few customers. As a result, LV circuits in North American-style networks can be modelled reasonably using service cable model templates (as in \cite{dugan2010ieee}). This is in stark contrast to European-style LV network models--for example, modelling just a single LV network resulted in a network model of over 4,500 nodes in \cite{gutierrez2019opf}. Additionally, the number of customers fed by a European-style MV primary circuit is often much larger, with over ten thousand customers fed by a single primary substation being a common occurrence.

Because of their size and data requirements (and DSOs reluctance to share potentially sensitive information), there are few readily accessible European MV--LV models. One approach that has been proposed to overcome the DSO data sensitivity issue is to build synthetic European test cases based on geographical data \cite{domingo2010reference}, with a similar approach taken in a US-based context in \cite{mateo2020building} and a Central American case study in \cite{valverde2017integration}. A set of non-synthetic, unbalanced LV circuits fed from a common MV primary substation are given in \cite{koirala2020non}. Other works use a `network allocation' approach, whereby LV networks are allocated to selected loads in an MV network \cite{escobar2020combined,deakin2020flexibility,gutierrez2019opf}. This approach is attractive as it allows for the use of thoroughly validated, real LV circuit models (it can be challenging to convert DSO geographic databases to clean electrical network models). For a more detailed review of distribution network models, we refer the interested reader to \cite{postigo2017review}.

To our knowledge, however, there are no full-scale unbalanced European style MV--LV networks that are openly available for researchers to work from. This view is mirrored by a recent review \cite{givisiez2020review}, which highlights the importance of a three-phase representation of MV--LV test circuits, whilst noting the scarcity of models of this type. The overhead involved with constructing and validating these types of circuits is non-trivial, and so this represents a significant gap.

In this work we present a pair of hybrid European unbalanced MV--LV models to address this gap. The models are neither fully synthetic nor fully physical, with over one hundred real LV feeders allocated to the MV loads. The circuits have both urban (underground) and rural (overhead) MV feeders, with the MV--LV construction allowing for the coupling between LV loads on adjacent networks to be captured. The models are available from
\begin{center}\small
\texttt{https://github.com/deakinmt/uk-mvlv-models}~,
\end{center}
and are designed to be a readily available test bed for the development of scalable controls that take advantage of the highly heterogeneous behaviour of LV circuits whilst accounting for MV level coupling.

The rest of this paper is structured as follows. Section \ref{s:2lvns} outlines both MV and LV circuit preprocessing stages and the load-network allocation steps, to clearly outline the assumptions made in the construction of the models. A number of validation steps are discussed in Section \ref{s:4val}, to check that the networks behave as expected. Some possible use cases of the models are presented in Section \ref{s:5case}, to highlight how these models could be used to study issues in the wider energy systems context. Salient conclusions are drawn in Section \ref{s:6conc}.

\section{MV--LV Circuit Pre-processing and\\Network Allocation}\label{s:2lvns}

In this work we use a hybrid approach for creating the MV--LV test cases, in which real LV networks are allocated to MV circuit models (as illustrated in Fig. \ref{f:xmplNtwks}). There are three steps in the creation of the models. First, MV and LV test circuits are preprocessed; then, the LV networks are allocated to the MV loads; finally, the MV--LV model is validated by studying load flow solutions under high and low demand conditions.

\subsection{LV Network Preprocessing}\label{s:2alv}
\begin{figure}
\centering
\subfloat[LVNS `Network 1' topology]{\includegraphics[width=0.36\textwidth]{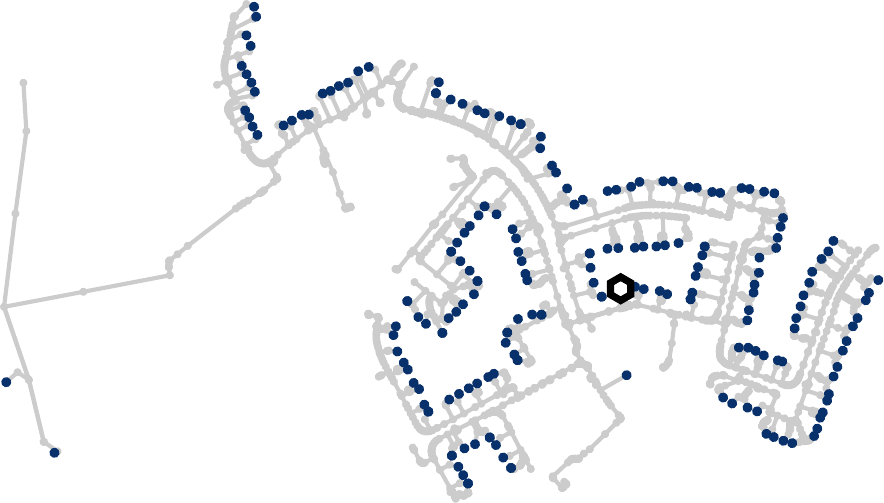}\label{f:plotNetwork_n1_n}}~\\
\subfloat[UKGDS `HV-UG' topology]{\includegraphics[width=0.42\textwidth]{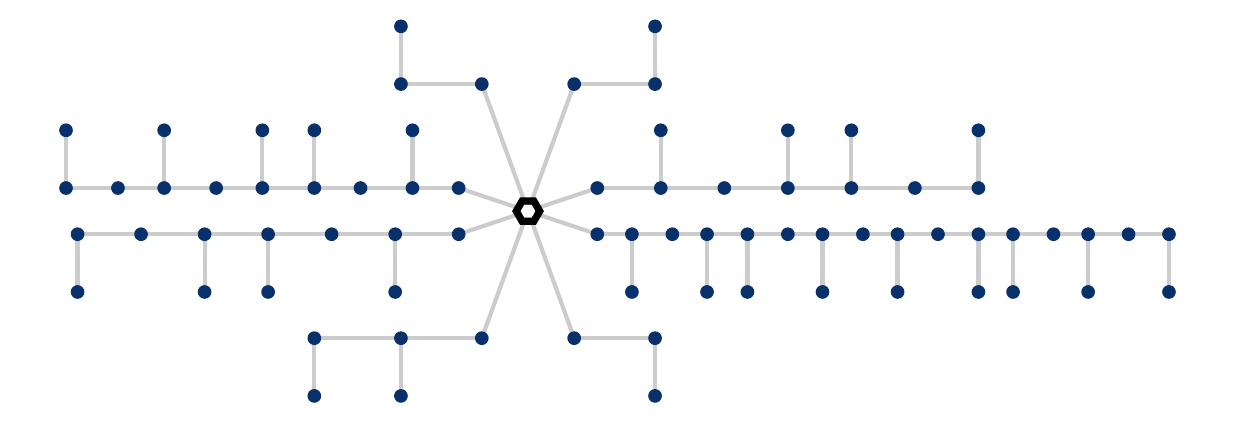}\label{f:plotNetwork_HV_UG_n}}~\\
\subfloat[Integrated MV--LV Models]{\includegraphics[width=0.36\textwidth]{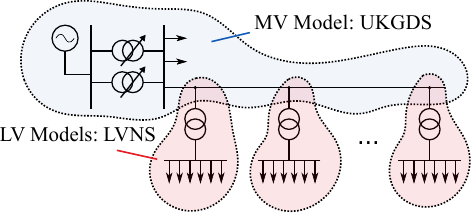}\label{f:mv-lv-modelling}}
\caption{The LVNS network models (e.g., (a)) are allocated to the loads of the UKGDS models (e.g., (b)) to create the integrated MV-LV models (c).}
\label{f:xmplNtwks}
\end{figure}

The Low Voltage Network Solution (LVNS) test circuits \cite{enwl2015low} were developed for studying the impacts of low-carbon technologies on European-style distribution networks. These circuits have been thoroughly validated and are well-documented (one of the feeders became the European 906-Node Low Voltage Test Feeder). There are 25 LV networks, disaggregated into 128 feeders. One of the networks, consisting of four individual feeders, is plotted in Fig. \ref{f:plotNetwork_n1_n}.

Three specific challenges were considered while preprocessing these circuits for use in the integrated MV--LV model. Firstly, there are a relatively large number of nodes per customer in the LVNS networks, which (due to the lack of modelled cable capacitance) can be spliced together to form individual branch models without changing the load flow solution \cite{rigoni2020open}. The steps taken to validate this step are described in the Appendix, with the number of nodes in each LV network being reduced by between 5 and 18 times.

Secondly, it is noted in the LVNS documentation that some feeders have unusually heavy unbalance \cite{espinosa2015low}. Therefore, individual LV feeders with more than 5\% zero sequence unbalance are removed, as are feeders explicitly mentioned as being unusually heavily unbalanced in \cite{espinosa2015low}. Additionally, as shall be noted in the next section, some LVNS networks are larger than the largest of all of the UK Generic Distribution System (UKGDS) MV loads. For example, `Network 17' has 883 loads which is (assuming a conservative 1 kW per load) much larger than the largest load of 436 kW on the UG circuit (Table \ref{t:tblMvInfo}). Therefore, in the three largest networks (networks 2, 15 and 17) one feeder is removed, so that the total number of loads is fewer than 500 in all circuits used for allocation. In total, 112 of the 128 feeders are available for allocation to the MV circuit.

\subsubsection{Transformer Sizing}
All transformers in the LVNS set are modelled as 800 kVA transformers, which in many cases represents an inefficient allocation of resource (the apparent overrating of these transformers has been noted in previous works such as \cite{koirala2020non}). Whilst there are reasons why they may be the true sizes in reality \cite{npg2018code}, following \cite{koirala2020non} it is assumed that these transformers are unlikely to be broadly representative of secondary substation ratings across European LV networks. 

Therefore, once any modifications have been made (in terms of excising unwanted feeders described previously) a transformer rating is specified according to after diversity maximum demand (ADMD) estimates from \cite{barteczko2015after} and then the transformer size is specified according to a UK-based network operator's domestic transformer sizing specifications. Specifically, the ADMD is calculated from the number of LV network loads $N_{\mathrm{Lds}}$ as
\begin{equation}\label{e:rated}
\mathrm{ADMD} = N_{\mathrm{Lds}}\times \max \left\{ S_{1} N_{\mathrm{Lds}}^{-\alpha}, S_{\infty} \right\}\,,
\end{equation}
where $S_{1}=5.325$ kVA is the single-load rating, $\alpha=0.262$ is the ADMD coefficient, and $S_{\infty}=1.5$ kVA is the minimum per-load ADMD \cite{barteczko2015after} (to ensure reasonable capacity is allocated when there are large numbers of loads). With this ADMD level, the smallest acceptable transformer rating was selected from \cite[Table 8]{npg2018code}. Transformer impedance values of 1 + $\jmath $5\% pu were assumed for the secondary transformers (with the pu base as the transformer rating), following \cite{ingram2003impact}.

\subsection{MV Network Preprocessing}
The UKGDS test systems define a set of meshed subtransmission (EHV) networks and a set of radial 11 kV MV models (referred to in the UKGDS documentation as `HV' models) \cite{foote2005defining}. These MV models are useful for the purposes described here as the circuits are of the correct voltage level and are of radial topology (the topology of one of the circuits is plotted in Fig. \ref{f:plotNetwork_HV_UG_n}).

From the seven MV UKGDS circuits, two circuits (the UG and UG/OH-A circuits) were chosen as the MV circuits for the test networks. These two circuits were selected because the sizes of the feeders are typical of UK-style circuits, but they show very different distributions of loads (see Table \ref{t:tblMvInfo}). In fact, the UG circuit has a demand 20\% greater than that of the UG/OH-A circuit but less than one quarter the number of LV circuits (Table \ref{t:tableCktsMv}).

\begin{table}
\centering
\caption{Two UKGDS circuits form the basis of the MV-LV Circuits, with up to eighteen thousand loads estimated (based on an assumption of 1.3 kVA per LV load \cite{ingram2003impact}).}\label{t:tblMvInfo}
\begin{tabular}{lllll}
\toprule
\multirow{2}{*}{\vspace{-0.4em}Ntwk. ID} & \multirow{2}{*}{\vspace{-0.4em}\begin{tabular}{@{}l@{}}Tot. Demand,\\ MVA\end{tabular}} & \multirow{2}{*}{\vspace{-0.4em}\begin{tabular}{@{}l@{}}MV\\Buses\end{tabular}} & \multirow{2}{*}{\vspace{-0.4em}\begin{tabular}{@{}l@{}}Est. LV\\ Loads\end{tabular}} & Load stats., kW \\
    \cmidrule(l{0.6em}r{0.9em}){5-5}
    & & & & (Min., Med., Max.) \\
    \midrule
UG & 25.3 & 78 & 18991 & (100, 344, 436) \\
UG/OH, A & 21.4 & 399 & 15727 & (4, 28, 420) \\
\bottomrule
\end{tabular}

\end{table}

As with the LV networks, we make modifications to the transformer impedance values to ensure the models are realistic. The models in the UKGDS MV networks have a per-unit transformer reactance between 2.5\% and 10\% on the transformer rating, whilst industry-approved simplified models \cite{ingram2003impact} and data from industry \cite{npg2019ltds} both show impedances closer to 20\% on the transformer rating. A more broadly representative value was therefore chosen by calculating the median resistance and reactance of all transformers from \cite{npg2019ltds} that had a low-voltage rating of 11 kV. This approach led to a per-unit impedance of 0.88 + $\jmath $19.97\% being chosen for primary substation transformers (using the transformer ratings as the base unit).

\subsection{MV Load--LV Network Allocation}

The LV Network allocation step was approached with the assumption that the peak loads on the MV network correspond closely to the designed peak load of LV customers, where each customer is assumed to have a demand of 1.3~kVA \cite{ingram2003impact}. To ensure that small loads would have a network allocated to them (the smallest LV network has 42 loads, whilst there are some small loads of just 4 kVA in the UG/OH circuit), the first feeder (`Feeder 1') from each of the LV networks was also considered for allocation.

The LV network allocation procedure consisted of two steps, with the goal of ensuring good coverage of LV networks across the MV--LV networks whilst ensuring MV branch power flows did not change too much.
\begin{itemize}
\item For each of the loads on the MV circuit, a random number was drawn from a normal distribution with unity mean and a standard deviation of 0.175;
\item Then, the network with a kVA rating closest to the multiplication of this random number and the MV load was allocated.
\end{itemize}
With this approach, all 25 of the full LV networks were picked between the two MV--LV circuits, as well as twenty of the individual LV feeders (for smaller loads). A summary of key circuit parameters is given in Table \ref{t:tableCktsMv}.

\begin{table}
\centering
\caption{Integrated MV-LV Circuit Parameters}\label{t:tableCktsMv}
\begin{tabular}{llll}
\toprule
Ckt. ID & No. Nodes & No. Lds. & No. LV Ckts. \\
\midrule
UG & 112,887 & 19,031 & 75 \\
UG/OH, A & 86,448 & 15,166 & 308 \\
\bottomrule
\end{tabular}

\end{table}

It is worth noting that the models in this work are focused on domestic-style LV networks, with no allocation of industrial or commercial load. As a result, there are likely to be more loads in Table~\ref{t:tableCktsMv} than there would be customers allocated to typical European-style primary substations of the power ratings given in Table~\ref{t:tblMvInfo}. Nevertheless, even with this taken into account there would still be a much larger number of loads than there are in, say, the IEEE 8500-node circuit, which has 1,177 individual loads modelled.

\section{Integrated MV--LV Model Validation}\label{s:4val}

The final stage in the model development was the model validation, ensuring that the circuits that have been built are reasonable in terms of the voltages and powers found at load flow solutions under realistic conditions. For this purpose, the load flow at a `High' demand and `Low' demand condition were considered, with individual LV customers having loads of 1.3 kVA and 0.16 kVA at unity power factor, based on \cite{ingram2003impact}.

\begin{figure}\centering
\subfloat[Voltages (MV)]{\includegraphics[width=0.16\textwidth]{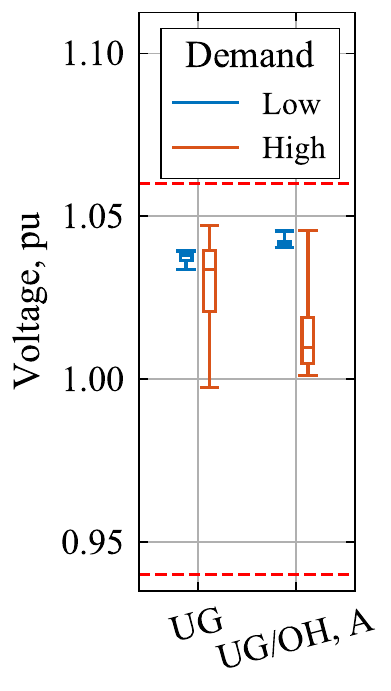}\label{f:pltMvLvRanges_MV}}
\subfloat[Voltages (LV)]{\includegraphics[width=0.16\textwidth]{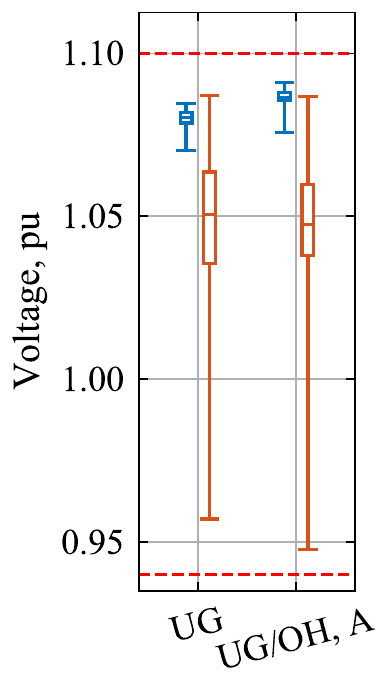}\label{f:pltMvLvRanges_LV}}
\subfloat[Powers]{\includegraphics[width=0.16\textwidth]{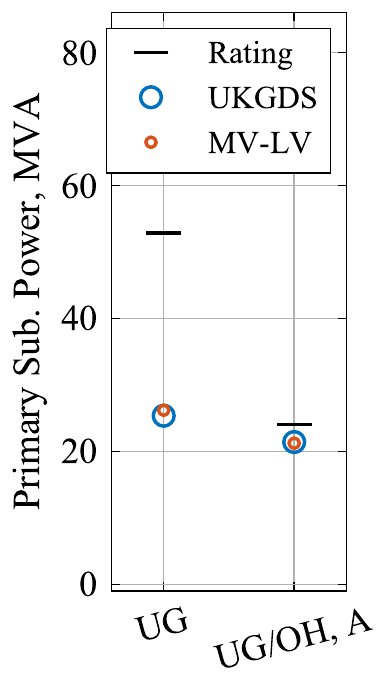}\label{f:pltCheckPwrsMv}}
\caption{Checking the MV and LV spread of voltages and the total powers of the two MV--LV integrated models. Voltages are calculated for both High and Low demand conditions (with 1.3 kW and 0.16 kW at all customers, respectively), whilst the substation powers are calculated during High demand conditions. The boxplots show the range, interquartile range and median voltages.}\label{f:pltMvLvRanges}
\end{figure}

The voltages under these conditions are plotted as boxplots in Figs. \ref{f:pltMvLvRanges_MV}, \ref{f:pltMvLvRanges_LV}. Both circuits tend to have voltages that are high during Low demand conditions, with a wide spread of voltages during High demand conditions. The load flow solution's voltages do not violate the UK statutory limits of (0.94, 1.06) pu for MV circuits and (0.94, 1.1) pu for LV circuits \cite{ena2017statutory}.

In addition, the powers seen at the primary substation are also calculated to ensure that the High demand condition does not violate thermal constraints, as plotted in Fig. \ref{f:pltCheckPwrsMv}. The LV network allocation has resulted in the MV--LV model powers remaining close to the powers of the UKGDS models from which they are taken. This is particularly important in the UG/OH case, for which the total combined rating of the primary substation transformers is very close to that of the UKGDS circuit's power.

\subsection{Disaggregated LV Network Voltages}

\begin{figure}\centering
\includegraphics[width=0.49\textwidth]{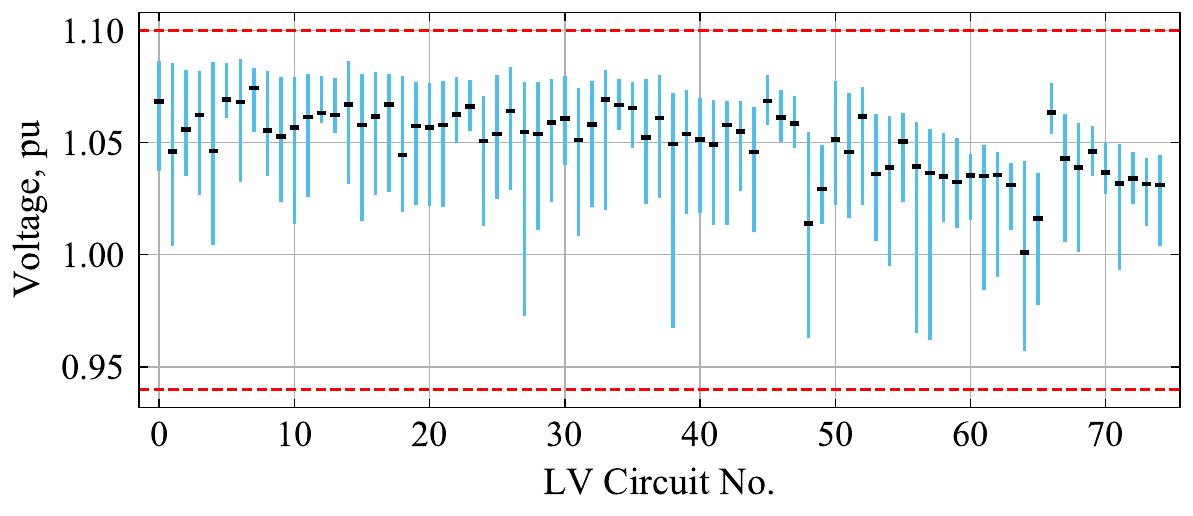}
\caption{The range and median voltages, represented as light blue lines and black dashes respectively, for each of the LV circuits allocated to the UG MV--LV circuit (under the High demand condition with 1.3 kW at each load).}
\label{f:pltMvLvVnom_90}
\end{figure}

\begin{figure*}\centering
  \includegraphics[width=0.96\textwidth]{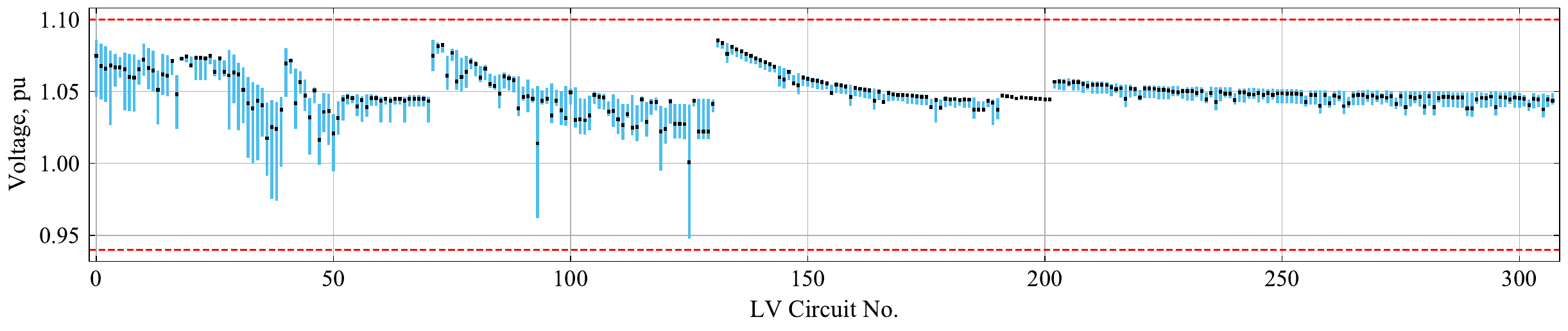}
  \caption{The range and median voltages, represented as light blue lines and black dashes respectively, for each of the LV circuits allocated to the UG/OH-A MV--LV circuit (under the High demand condition with 1.3 kW at each load).}
  \label{f:pltMvLvVnom_91}
\end{figure*}

The MV models from the UKGDS networks are operated radially, with a number of feeders coming from the primary substation--for example, there are eight MV feeders on the UG circuit, as shown in Fig. \ref{f:plotNetwork_HV_UG_n}. This structure becomes apparent in the solution if the voltages on the LV circuits are plotted against the LV circuit number (so that LV circuits on the same MV feeder are adjacent), as shown in Figs. \ref{f:pltMvLvVnom_90}, \ref{f:pltMvLvVnom_91}. This is particularly noticeable on the UG/OH-A solution, with voltage coupling clearly visible on the long overhead MV feeders. It is very important to note, however, that the voltage response of LV circuits is very heterogeneous--most of the loads are well insulated from large voltage changes.

There are some qualitative differences in the solutions (Figs. \ref{f:pltMvLvVnom_90}, \ref{f:pltMvLvVnom_91}) -- there are many more LV circuits in the UG/OH model than in the UG model, but less variation in solutions. This is because the smaller feeders within the LVNS test cases tend to have a small voltage drop compared to the larger, urban feeders (large feeders are much more prevalent in the UG model).

\subsection{Comparison: Real LV voltage measurements}

\begin{figure}\centering
\subfloat[Voltage measurement boxplots]{\includegraphics[width=0.23\textwidth]{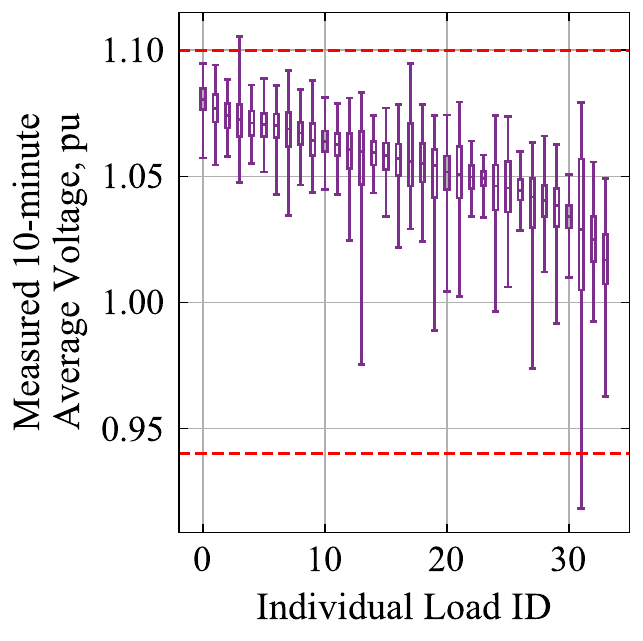}\label{f:pltMixVolt_2020_01}}
~
\subfloat[Voltage QQ plot]{\includegraphics[width=0.24\textwidth]{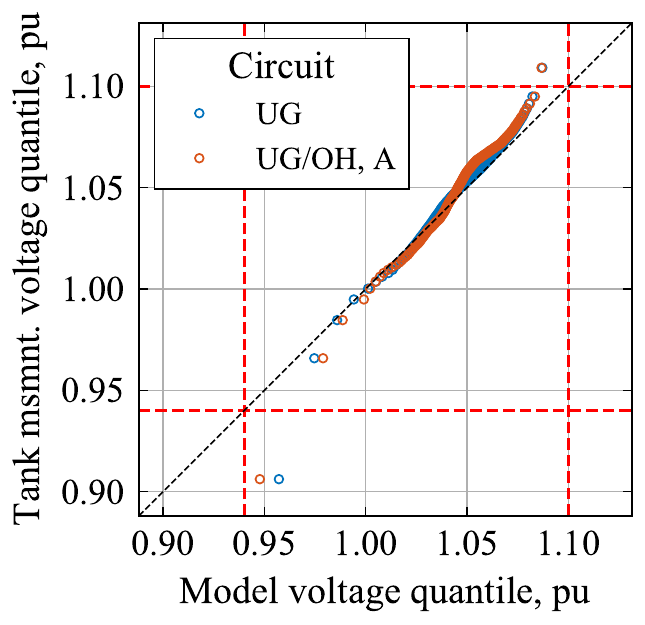}\label{f:pltMixVoltQq_2020_01}}
\caption{10-minute average voltage measurements were collected at 104 smart hot water tanks from 5--8pm each day of January 2020. The boxplots (a) show the range, interquartile range and the median voltages for a subset of these tanks. The quantile--quantile (QQ) plot (b) shows that the model fits to the measured voltage data well, except in the tails of the distributions.}
\label{f:pltMixVolt}
\end{figure}

To study both the validity of the model and consider if the aforementioned heterogeneity in voltages exists in practice, real voltage measurements were obtained from 104 instrumented UK-based hybrid smart hot water tanks. The tanks cover a wide geographic area and so are assumed to be broadly representative of LV customer voltages. Boxplots of voltage data\footnote{The tanks primarily use gas for heating; occasional periods when the electric heating element was used instead the voltage measurements were discarded. This was to avoid bias due to voltage drops within a customer's property beyond the meter, which is not considered within LV network models.} for the times of 5--8pm thoughout January 2020 are shown in Fig. \ref{f:pltMixVolt_2020_01}, and the quantile--quantile (QQ) plot of Fig. \ref{f:pltMixVoltQq_2020_01} compares the distribution of voltage measurements during this period against the voltages under the High demand condition for both circuits. 

The QQ plot shows that the model approximates the measured data well. There is some disagreement in the low voltage tail of the distribution, which is not unexpected as the MV--LV models only consider uniform demand, where real LV demands can show large variation about the mean. Nevertheless, the measurements corroborate the observation that whilst most customers do not have voltage issues, a significant fraction of loads do show occasional voltage violations.

\section{Flexibility Modelling Case Studies}\label{s:5case}

The models are intended to be a sandbox for the design of algorithms, with many possible applications. Here we present two short case studies, based on (i) a peer--peer energy trading platform within a single MV--LV network, and (ii) domestic smart hot water tanks responding to a TSO--DSO signal to avoid curtailment of variable renewable generation.

\subsection{Peer--peer Energy Trading and Constraints}
The first case we consider is that of a peer--peer trading platform, whereby a set of customers can form a coalition and trade energy between themselves in a relatively small geographic area to reduce their network bills. The end goal of such a scheme is to empower consumers and incentivise trading and/or sharing of resources that could otherwise be under-utilised \cite{morstyn2018using}.

\begin{figure}\centering
\includegraphics[width=0.48\textwidth]{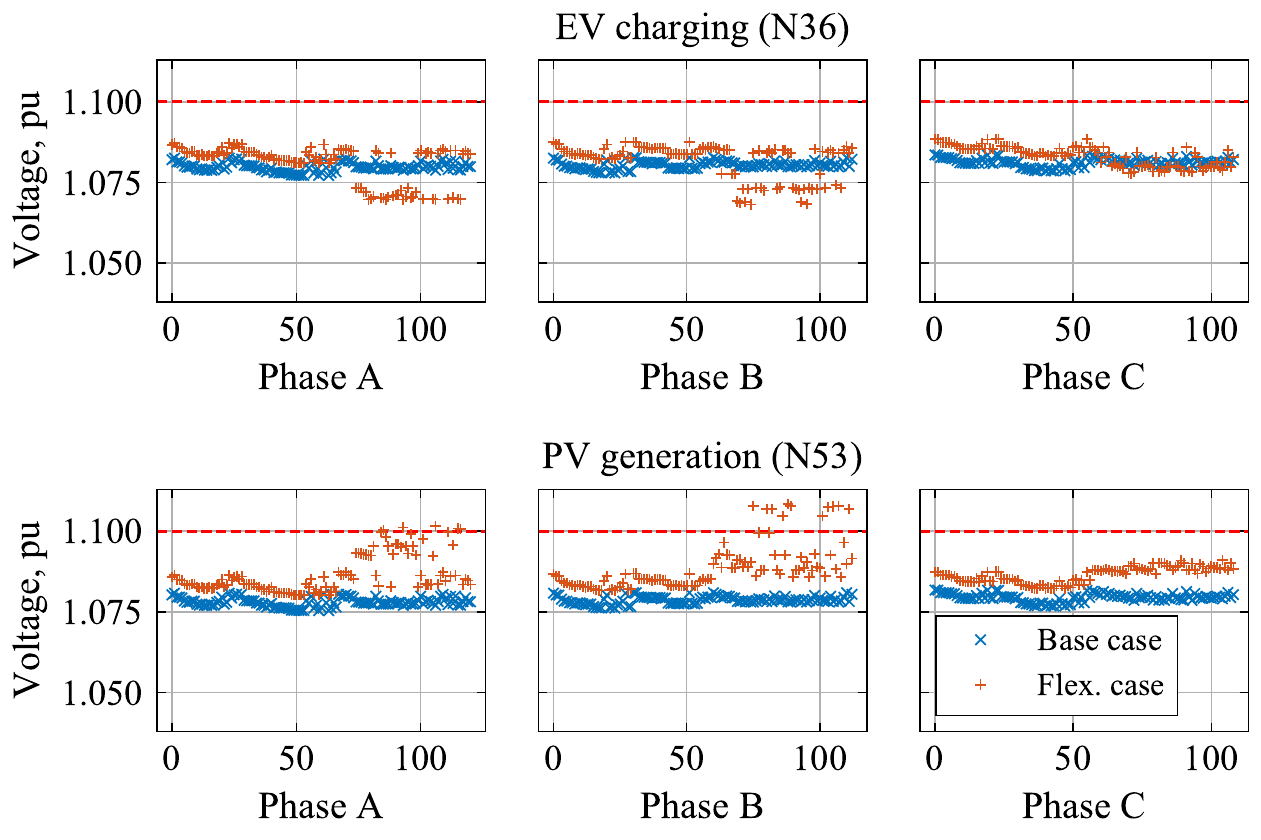}
\caption{The voltages for Network 36 (N36) and Network 53 (N53) of the UG MV--LV model when 3 kW of power is `sent' from the PV generation from Network 53 to the EVs in Network 36 via a peer--peer contract. This results in overvoltages on Phase A and B of Network 53.}\label{f:pltCaseStudyC}
\end{figure}

In this first scenario we envision trading from one LV network to an adjacent LV network, with a set of fifteen 3~kW solar PV generators agreeing in advance to power a set of fifteen 3~kW electric vehicle (EV) chargers during the Low demand condition. In this instance, the solar PV cannot generate without causing over-voltages (Fig. \ref{f:pltCaseStudyC}). Therefore, the peer--peer platform will have to instruct the PV generators to reduce the power which is being traded, resulting in a reduction in income from the generator's asset and an increase in the power price for the EV demands. If this issue could not be cleared with normal DSO actions, regulations would be required to stipulate who pays the opportunity cost, i.e., whether it is the platform participants who are affected financially, or if the DSO must provide compensation to those customers.

\subsection{System-Wide Downward Flexibility}
The second case we consider is that of system-wide flexibility provision, with the assumption that some customers have access to smart hot water tanks that are responsive to price signals. If there is transmission level congestion, a TSO could signal for increased demand, in which case the signal would be indiscriminate across the whole MV--LV circuit. 

\begin{figure}\centering
\includegraphics[width=0.495\textwidth]{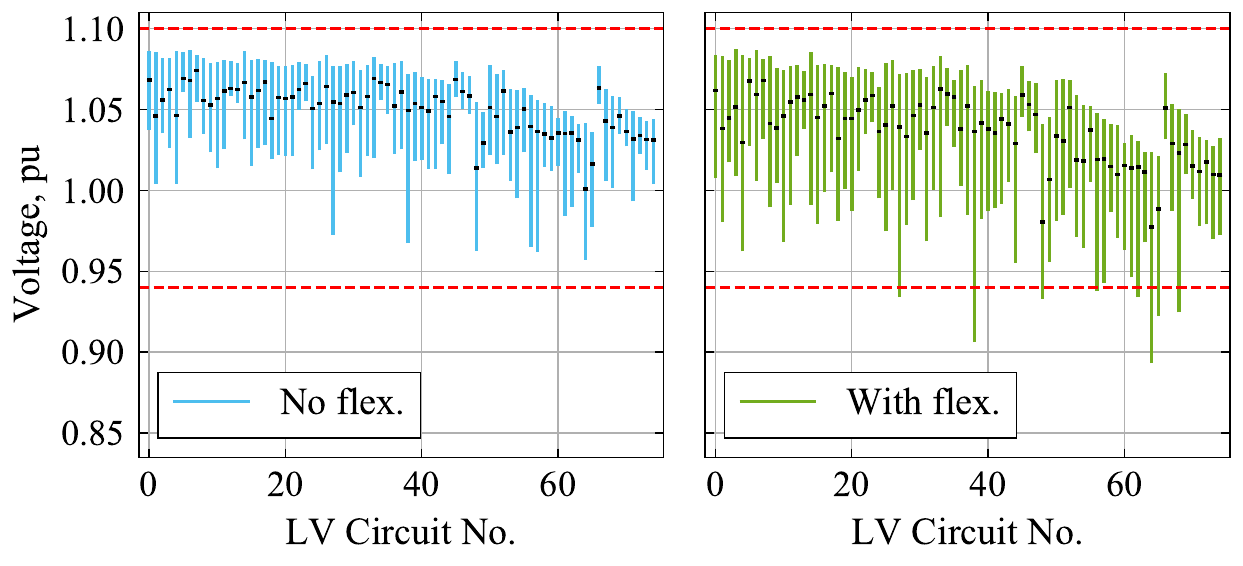}
\caption{The range and median voltages seen on each of the UG circuit's LV networks, for the case with no flexibility, and the case with flexibility enabled (with 13\% of customers increasing their demand by 3 kW).}\label{f:pltCaseStudyA}
\end{figure}

As an example of a possible response to this signal, the demand is increased by 30\%, effected by increasing the demand of 13\% of locations by 3 kW (locations for demand response are selected at random). The voltages before and after the demands are increased are plotted in Fig. \ref{f:pltCaseStudyA}. This illustrates that there are a small number of customers that would experience undervoltages. In the UK, it would depend on the frequency of these violations for a DSO to determine if they are required to be cleared--DSOs can permit the voltage to drop to 0.85 pu up to 5\% of the time \cite{ena2017statutory}. 

\subsection{Discussion: MV--LV in the context of Whole Energy Systems}\label{s:5discussion}

Historically, modelling of MV--LV systems has often made assumptions about almost-homogeneous networks, as in \cite{ingram2003impact}. In contrast, the whole energy systems vision will result in already diverse distribution networks interacting with other energy vectors and information systems to enable future prosumers to decarbonise energy systems at pace.

For network planners to get feedback from models to understand how they respond to varying design choices, the time spent evaluating network performance should ideally be a few seconds or less (see, e.g., \cite[Ch. 5]{nielsen1994usability}). On the other hand, automatic control systems may require on-line decision-making in a fraction of a second. Without efficient data analytics and visualisation, the deployment of flexibility services and smart controls risks either cautiously under-utilising the network, or aggressively providing flexibility services at the expense of system security and quality of supply.

\section{Conclusions}\label{s:6conc}
This paper has presented a pair of network models, designed to be a testbed for the development of algorithms for studying unbalanced European-style MV--LV systems. The networks have been built by allocating real LV networks to two widely-used MV circuit models. It is hoped that the provision of these models will reduce the barriers required for researchers and engineers to access and study these types of networks. The scale of the networks is a particular challenge, with the number of nodes ten times that of the largest IEEE unbalanced distribution test feeder. 

As energy systems' geographic and temporal coupling increases, efficient computational techniques become progressively more important for the understanding and optimization of network operations and planning. The development of these models has demonstrated the heterogeneous nature of MV--LV systems, corroborated by measured LV customer voltage data. It is concluded that only by embracing this diversity that effective planning and operational decisions can be made that will enable decarbonisation of the whole energy system in a cost-effective and equitable way.

\section*{Acknowledgements}

The authors are grateful for the voltage measurement data from Mixergy Ltd., for helpful feedback on the networks from Jialiang Yi of UK Power Networks, and for helpful discussions and model testing from Ilias Sarantakos, Ridoy Das and Myriam Neaimeh of Newcastle University.

\section*{Appendix: LVNS LV Network Splicing}

\begin{figure}\centering
\includegraphics[width=0.48\textwidth]{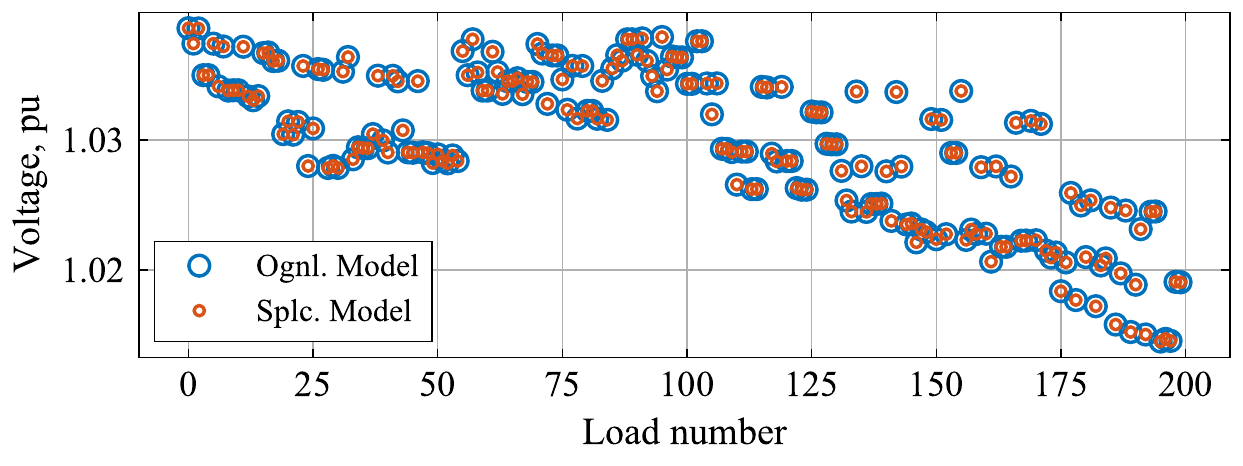}
\caption{The relative voltage error \eqref{e:errV} for LVNS Network 1 has value $\epsilon_{V}=4.22\times 10^{-12}$. The full model (Ognl.) has 9444 nodes whilst the spliced model (Splc.) has just 1269 nodes.}\label{f:validate}
\end{figure}

To demonstrate the accuracy of the spliced models (Section \ref{s:2alv}), the relative error in voltages and losses $\epsilon_{V},\,\epsilon_{\mathrm{Loss}}$ (respectively) are calculated as
\begin{equation}\label{e:errV}
\epsilon_{V} = \dfrac{\| V_{\mathrm{Lds}}^{\mathrm{Ognl.}} - V_{\mathrm{Lds}}^{\mathrm{Splc.}}\|}{\|V_{\mathrm{Lds}}^{\mathrm{Ognl.}}\|}\,, \quad  \epsilon_{\mathrm{Loss}} = \dfrac{|P_{\mathrm{Loss}}^{\mathrm{Ognl.}} - P_{\mathrm{Loss}}^{\mathrm{Splc.}} |}{|P_{\mathrm{Loss}}^{\mathrm{Ognl.}}|}\,,
\end{equation}
where $V_{\mathrm{Lds}}^{(\cdot )}$ are the complex voltages at each of the loads, $P_{\mathrm{Loss}}^{(\cdot)}$ are the total feeder losses, and superscript $(\cdot)^{\mathrm{Ognl}},\,(\cdot)^{\mathrm{Splc}}$ represent the original (full) model and the spliced model, respectively. The solution for Network 1 is shown in Fig. \ref{f:validate}. The value of $\epsilon_{V}$ is no greater than 3$\times 10^{-11}$ for all 25 of the networks, whilst the value of $\epsilon_{\mathrm{Loss}}$ was no greater than $4\times 10^{-9}$. The networks were reduced in size by between five and eighteen times, with the maximum network size being reduced from 55,536 nodes to 5220 nodes.


%
%
%
%
%
\bibliography{refs}{}
\bibliographystyle{IEEEtran}

\end{document}